\documentclass[a4paper,prd,preprint,amsmath,
               superscriptaddress,nofootinbib,floatfix,showpacs]{revtex4}
\usepackage{graphicx}
\newcommand{\bea}{\begin{eqnarray}}
\newcommand{\ena}{\end{eqnarray}}

\renewcommand{\a}{\alpha}
\renewcommand{\b}{\beta}

\newcommand{\e}{\epsilon}

\begin{document}
\title{The renormalization group effect to the bi-maximal mixing }
\date{8/10/2003}
\author{Takahiro Miura}
\email{miura@het.phys.sci.osaka-u.ac.jp}
\author{Tetsuo Shindou}
\email{shindou@het.phys.sci.osaka-u.ac.jp}
\author{Eiichi Takasugi}
\email{takasugi@het.phys.sci.osaka-u.ac.jp}
\affiliation{Department of Physics, Graduate School of Science,
		Osaka University Toyonaka, Osaka 560-0043, Japan}
\begin{abstract}
We discuss whether the bi-maximal mixing given at GUT is 
consistent with the solar mixing at the normal side at the 
low energy. We consider the radiative corrections due to 
the $\tau$--Yukawa, the neutrino--Yukawa and the slepton 
threshold corrections and discuss in what situation the 
maximal solar angle rotates towards the normal side. In this 
scheme, the $|V_{13}|$ and the Dirac CP phase $\delta$ are 
induced radiatively and we estimate these values.  
\end{abstract}
\preprint{OU-HET 454}
\preprint{hep-ph/0308109}

\maketitle

\section{Introduction}
In view of the data by Super-K \cite{SK-Solar}, SNO \cite{SNO-Solar}, 
the solar mixing angle will not be maximal and its best 
fit value is
\bea
\tan^2 \theta_\odot\simeq 0.34\;,
\ena
and the solar mass squared difference is precisely determined 
by the KamLAND data \cite{KamLAND} as
\bea
\Delta m_\odot^2\equiv m_2^2-m_1^2 
\simeq 6.9\times 10^{-5}\;{\rm eV}^2\;. 
\ena

On the other hand, the bi-maximal mixing scheme \cite{BiMax} of the neutrino 
mixing is quite attractive theoretically, but it predicts 
$\tan^2 \theta_\odot=1$, which contradicts with the 
experimental data. However, if we consider that the bi-maximal 
mixing is derived by GUT, we have know the solar mixing angle 
at low energy where the neutrino oscillation experiments are 
preformed. This can be made by analyzing the 
renormalization group effect to the neutrino mass matrix.  

In our previous paper \cite{mst1}, we analyzed the effect of the 
renormalization group due to $\tau$-Yukawa coupling 
to the neutrino mass derived from the see-saw mechanism, 
and we found that the solar angle increases as the 
energy scale decreases. That is, if $\tan^2 \theta_\odot=1$ 
at GUT scale, $M_X$, then the solar angle moves towards the dark-side, 
$\tan^2 \theta_\odot >1$ at low energy, $m_Z$, as the energy scale 
moves from  $M_X$ to $m_Z$. This may be a serious trouble 
for the bi-maximal mixing scenario at GUT energy. 

In this paper, we discuss the question whether we can achieve 
the normal-side solar mixing angle, $\tan^2 \theta_\odot <1$, 
at low energy by quantum effects, starting form the maximal 
mixing at GUT energy. It is known \cite{Stability} that 
the quantum effects to the mixing angles is small when the 
neutrino mass spectrum is hierarchical. Therefore, we are 
forced to consider the quasi-degenerate neutrino scenario, 
if we seek this possibility.

\section{The renormalization effect to the bi-maximal mixing} 
We take the basis where the charged lepton mass matrix is 
diagonal.  The neutrino mass matrix 
which gives to the bi-maximal mixing at $M_X$ is given by 
\bea
m_{\nu}(M_X)=O_BD_{\nu}O_B^T\;,
\ena
with
\bea
O_B=\begin{pmatrix}
\frac{1}{\sqrt{2}}&-\frac{1}{\sqrt{2}}&0\cr
\frac{1}{2}&\frac{1}{2}&-\frac{1}{\sqrt{2}}\cr
\frac{1}{2}&\frac{1}{2}&\frac{1}{\sqrt{2}}\cr\end{pmatrix}\;,
\ena
and
\bea
D_{\nu}=\mathrm{diag}(M_1,M_2e^{i\alpha_0},M_3e^{i\beta_0})\;.
\ena
where  $M_i$ are neutrino masses at GUT scale and taken to be 
real positive and almost degenerate. The phases  
$\alpha$ and $\beta$ are CP violating Majorana phases \cite{Majoranaphase}. 
This mass matrix contains five parameters.  

We assume that the neutrino mass matrix at the low energy, 
$m_{\nu}(m_Z)$ is related to that of GUT scale, $m_{\nu}(M_G)$ by   
\bea
m_{\nu}(m_Z)=m_{\nu}(M_G)+K
 m_{\nu}(M_G)+m_{\nu}(M_G)K\;.
\ena
As we shall discuss in detail later, we parametrize $K$ as
\bea
K=\begin{pmatrix}\e_e & 0 & 0\cr 0&0&0 \cr 0&0&\e_\tau \cr\end{pmatrix}\;.
\ena

There are several origins for $\e_e$ and $\e_\tau$, 
the effect due to the $\tau$-Yukawa coupling \cite{tau_yukawa}, 
the effect due to the neutrino-Yukawa couplings 
\cite{nu_yukawa_1, nu_yukawa_2}, 
the threshold correction due to heavy right-handed neutrino 
mass splitting \cite{nu_yukawa_1, nu_yukawa_2} 
and the slepton mass splitting \cite{slepton}, which we 
discuss in the next section. 

In below, we assume that $\e_e$ and $\e_\tau$ are small quantities 
of order $10^{-3}$ and consider the quasi-degenerate mass case, 
\bea
M_1&\sim & M_2 \sim M_3\;,\nonumber\\
|\epsilon_e|&\sim &|\epsilon_\tau|<<1\;,\nonumber\\
|\Delta_{31}|&>>&|\Delta_{21}|\sim \epsilon M_i. 
\ena
where 
\bea
\Delta_{21}=M_2^2-M_1^2\;,\; \Delta_{31}=M_3^2-M_1^2\;.
\ena
Now, we discuss the effect of $K$ 
to the neutrino mass matrix in the first order approximation 
of $\e_e$ and $\e_\tau$. 

We transform $m_\nu(m_Z)$ by $O_B$ as
\bea
\bar m_\nu(m_Z)&=&O_B^T m_\nu(m_Z)O_B \nonumber\\
   &=&D_\nu+O_B^TKO_B D_\nu+ D_\nu O_B^TKO_B\;,
\ena
where
\bea
O_B^TKO_B = \frac{\epsilon_\tau}4
\begin{pmatrix}
1&1&\sqrt{2}\cr
1&1&\sqrt{2}\cr
\sqrt{2}&\sqrt{2}&2\cr\end{pmatrix}
+\frac{\epsilon_e}2 \begin{pmatrix}
1&-1&0\cr
-1&1&0\cr
0&0&0\cr\end{pmatrix}\;.
\ena

Since $\bar{m}_{\nu}$ is a complex symmetric matrix, we consider 
the hermitian quantity $\bar{m}_{\nu}^{\dagger}\bar{m}_{\nu}$ which 
is given by
\bea
\bar{m}_{\nu}^{\dagger}\bar{m}_{\nu}\simeq
 (1+2\epsilon_e+\epsilon_\tau)M_1^2+Y_0+Y_{e}+Y_{\tau}
\ena
where elements of $Y_i$ are given by 
\bea
Y_0&=&{\rm diag}(0,\Delta_{21},\Delta_{31})\;,
\nonumber\\
Y_{e}&\simeq &-2\e_e
\begin{pmatrix}
0 &1+e^{i\alpha_0}&0\cr
1+e^{-i\alpha_0}&0&0\cr
0&0&0\cr\end{pmatrix}M_1^2
\;,\nonumber\\
Y_{\tau}&\simeq &\e_\tau \begin{pmatrix}0&1+ e^{i\a_0}&
\frac{1+R^2+2Re^{i\b_0}}{2\sqrt 2} \cr
 1+ e^{-i\a_0} &0 & \frac{1+R^2+2R e^{i(\b_0-\a_0)}}{2\sqrt 2}\cr
 \frac{1+R^2+2Ee^{-i\b_0}}{2\sqrt 2} &
 \frac{1+R^2+2R e^{-i(\b_0-\a_0)}}{2\sqrt 2}&
 2R^2-1\cr\end{pmatrix}M_1^2\;,
\ena
where $R=M_3/M_1$ and 
we neglected terms of order $\epsilon_i \Delta_{21}$. 
We observe that  
$\Delta_{31}$ dominates over all other terms in the 
matrix $Y_0+Y_{e}+Y_{\tau}$, 
so that we can employ the see-saw calculation. 
By the unitary matrix 
\bea
V_{3} \simeq \begin{pmatrix}1&0&
        \frac{(Y_{\tau})_{13}}{(Y_{\tau})_{33}} \cr
    0&1&\frac{(Y_{\tau})_{23}}{(Y_{\tau})_{33}}\cr
    -\frac{(Y_{\tau})_{13}^*}{(Y_{\tau})_{33}}    &
    -\frac{(Y_{\tau})_{23}^*}{(Y_{\tau})_{33}}   &1\cr\end{pmatrix}\;,
\ena
where $Y_0+Y_{e}+Y_{\tau}$ is block diagonalized 
in the approximation to neglect 
terms $(Y_{\tau})_{3i}^*(Y_{\tau})_{j3}/(Y_{\tau})_{33}$, $(i,j\neq 3)$ 
which are of order 
$\sim \epsilon^2 M_i^2M_j^2/\Delta_{31}$. 
We diagonalize the remaining $2 \times 2$ matrix by 
\bea
V_{2}=
\begin{pmatrix}
c&-se^{i\alpha_0/2}&0\cr
se^{-i\alpha_0/2}&c&0\cr
0&0&1\cr\end{pmatrix}\;,
\ena
with $c=\cos\theta$ and $s=\sin\theta$, such that  
\bea
V_{2}^{\dagger}V_3^\dagger \bar{m}_{\nu}^{\dagger}\bar{m}_{\nu}V_3 V_{2}
=\mathrm{diag}(m_1^2,m_2^2,m_3^3)\;.
\ena
We find
\bea
m_i&\simeq &M_i\;,\nonumber\\
\Delta m^2_{\rm atm}&=&m_3^2-m_1^2 \simeq \Delta_{31}\;,
\nonumber\\ 
\Delta m^2_{\odot}&=&m_2^2-m_1^2\simeq \frac{\Delta_{21}}{\cos
2\theta}\;,
\ena
and the angle $\theta$ is expressed by replacing $M_i$ with $m_i$ as 
\bea
\sin2\theta
=-\frac{2(-2\e_e+\e_\tau)m_1^2\cos\frac{\alpha_0}{2}}
{\Delta m_{\odot}^2}\;.
\ena

By applying the transformation to $m_\nu(m_Z)$, we find
\bea
(O_BV_3V_{2})^Tm_{\nu}(m_Z)O_BV_3V_{2}\simeq
\mathrm{diag}(m_1,m_2e^{i\alpha_0},m_3e^{i\beta_0})\;,
\ena
with $m_i>0$, so that Majorana phases are determined. 
Thus, the mixing matrix which diagonalizes $m_\nu(m_Z)$ is 
$V=OV_3V_{12}P$ with 
$P={\rm diag}(1,e^{-i\alpha_0/2},e^{-i\beta_0/2})$. 
The neutrino mixing matrix is parameterized by
\bea
V=V_{MNS}P_{M}\;,
\ena
where $V_{MNS}$ is the MNS matrix \cite{msn} and 
$P_{M}$ is the Majorana phase matrix
\bea
P_{M}={\rm diag}(1,e^{i\a_M},e^{i\b_M})\;,
\ena
which represents the Majorana CP phases \cite{Majoranaphase}. 
From the above information, we find
\bea
V_{MNS}=\begin{pmatrix}
c_\odot &-s_\odot &{\rm sgn}(\e_\tau)|V_{13}|e^{-i\delta}\cr
\frac{s_\odot}{\sqrt 2}&\frac{c_\odot}{\sqrt 2} &-\frac1{\sqrt 2}\cr
\frac{s_\odot}{\sqrt 2}&\frac{c_\odot}{\sqrt 2} & \frac1{\sqrt 2}\cr
\end{pmatrix}
\;,
\ena
where 
\bea
s_{\odot}=\frac1{\sqrt 2}|c+ s e^{i\frac{\a_0}{2}}|\;,
\qquad 
c_{\odot}=\frac1{\sqrt 2}| c- se^{-i\frac{\a_0}{2}}|\;,
\ena
and the induced parameters, $|V_{13}|$ and the Dirac phase $\delta$ 
are given by
\bea
|V_{13}|&=&\frac{|\e_\tau| m_1m_3 
\sin\frac{\alpha_0}{2}}
{\Delta m^2_{\rm{atm}}}\;,\\
\delta&=&\xi_1+\xi_2+\frac{\alpha_0}2-\frac{\pi}2-\beta_0\;,
\ena
where $\xi_1={\rm arg}(c-se^{-i\frac{\a_0}{2}})$ and 
$\xi_2={\rm arg}(c+se^{i\frac{\a_0}{2}})$. 
The Majorana CP violation phases are given by
\bea
\alpha_M=\xi_2-\xi_1-\frac{\alpha_0}2\;,\qquad
\beta_M=\xi_2-\frac{\beta_0}2\;.
\ena

It may be worthwhile to mention that the effect by $K$ in Eq.(6) 
is to rotate the solar angle, to change the solar mass-squared 
difference, to induce $|V_{13}|$ and $\delta$. This effect is 
possible only for the quasi-degenerate neutrino mass scenario.

\section{The effect to the mixing matrix}
We found that the renormalization group affects to 
$\theta_\odot$ and $\Delta m_\odot^2$, and 
induces $|V_{13}|$ and $\delta$. In the following, we examine 
these quantities in detail.  

\begin{enumerate}
\item[(a)] The solar mixing angle

From Eq.(23), we find
\bea
\tan^2\theta_{\odot}=
\frac{1+ \sin 2\theta \cos \frac{\a_0}2}
{1- \sin 2\theta \cos \frac{\a_0}2}
\;,
\ena
so that $\tan^2\theta_{\odot}$ moves towards the normal side 
if $\sin 2\theta \cos(\a_0/2)=-\cos 2\theta_\odot <0$ and to the dark
side 
if $\sin 2\theta \cos(\a_0/2)=-\cos 2\theta_\odot >0$. 
From Eqs.(18), we find 
\bea
\sin2\theta\cos(\frac{\a_0}{2})
=-2(-2\e_e+\e_\tau)\cos^2(\frac{\alpha_0}{2})\frac{m_1^2}{
\Delta m_{\odot}^2}\;.
\ena
Thus,  we conclude for $\Delta m_{\odot}^2>0$
\bea
\tan^2\theta_{\odot}=\left \{\begin{matrix}{\rm normal \; side}
  \qquad -2\e_e+\e_\tau>0\cr
  {\rm dark \; side} \qquad -2\e_e+\e_\tau<0 \cr\end{matrix}
\right.
\ena

Now, we consider the allowed range of $\a_0/2$. From 
$\sin 2\theta =-\cos 2\theta_\odot/\cos(\a_0/2)$, 
\bea
|\cos \frac{\a_0}2|<\cos 2\theta_\odot\;.
\ena
By eliminating $\sin 2\theta$, we find
\bea
(-2\e_e+\e_\tau)=
\frac{ \cos 2\theta_{\odot}}2
\frac1{\cos^2\alpha_0/2} \frac{\Delta m_{\odot}^2}
{m_1^2}\;.
\ena
Numerically, we obtain
\bea
-2\e_e+\e_\tau = 7.2\times 10^{-3}
\left(\frac{0.49}{\cos (\a_0/2)}\right)^2
\left(\frac{0.1{\rm eV}}{m_1}\right)^2
\left(\frac{\Delta m_{\odot}^2}{6.9\times 10^{-5}{\rm eV}^2}\right)
\;,
\ena
where we used $\cos 2\theta_{\odot}=0.49$. 
The value of $-2\e_e+\e_\tau$ should be smaller than $7.2\times 10^{-3}$ 
for $m_1=0.1$eV, 
because $|\cos\frac{\alpha_0}{2}|$ must be larger than 
$\cos 2\theta_{\odot}$. 
Therefore, typically we need $\e_i$ of order $5\times 10^{-3}$ in order 
to move $\tan^2 \theta_\odot$ from 1 to 0.43. $-2\e_2+\e_\tau$ is 
proportional to $1/m_1^2$, 
the larger $\e_i$ is required if $m_i\simeq m$ is smaller than 0.2eV. 

\item[(b)] The size of the induced $|V_{13}|$

From Eq.(24), we find
\bea
|V_{13}|=0.018\left(\frac{|\e_\tau|}{5\times 10^{-3}} \right)
\left(\frac{m_1m_3}{(0.1)^2{\rm eV}^2}\right) 
\left(\frac{2.5\times 10^{-3}{\rm eV}^2}{\Delta m^2_{\rm{atm}}}\right)
\left(\frac{\sin\frac{\alpha_0}{2}}{0.87}\right)
\;,
\ena
The induced $|V_{13}|$ is small if the neutrino masses are 
less than 0.1 eV and if $\e_\tau$ is of order $5\times 10^{-3}$. 
The larger $V_{13}|$ is expected when both $\e_e$ and $\e_\tau$ are 
of order of $10^{-2}$ and the neutrino masses are about 0.2eV.

\item[(c)] The size of the induced $\delta$

We consider $\a_0/2$ in the region of $0<\a_0<2\pi$. 
From the relation, $\sin 2\theta \cos\frac{\a_0}2=-\cos
2\theta_\odot<0$, 
we define
\bea
\sin 2\theta =-x\;\;,\;\;x=\frac{\cos
2\theta_\odot}{\cos\frac{\a_0}2}\;.
\ena
Then, the allowed region of $x$ is 
\bea
\cos 2\theta_\odot <|x|<1\;,\;\; \left|\frac{\a_0}2 
\right|<2\theta_\odot\;. 
\ena 
We find two cases, 
\bea
\tan \xi_1&=&- \frac{\sqrt{1-\left (\frac{\cos
2\theta_\odot}{|x|}\right)^2}}
{\left(\frac{\sqrt{1+|x|}+
\sqrt{1-|x|}}{\sqrt{1+|x|}-\sqrt{1-|x|}}\right)
+\frac{\cos 2\theta_\odot}{|x|}}\;\;,\nonumber\\
\tan \xi_2&=&-\frac{\sqrt{1-\left (\frac{\cos
2\theta_\odot}{|x|}\right)^2}}
{\left(\frac{\sqrt{1+|x|}+\sqrt{1-|x|}}{\sqrt{1+|x|}-\sqrt{1-|x|}}\right)
-\frac{\cos 2\theta_\odot}{|x|}}\;\;.
\ena
Here, we consider $\cos \a_0/2>0$ case. 
For $x\to 1$, i.e., $\a_0/2\to 2\theta_\odot$, we find for both cases 
$\xi_1+\xi_2=-\pi/2$ and thus 
$\delta=2\theta_\odot-\pi-\beta_0\sim -(2/3)\pi-\b_0$. 
For $x \to \cos 2\theta_\odot$, i.e., $\a_0 \to 0$, we find 
$\xi_1=\xi_2=0$ so that $\delta=-\pi/2-\b_0$. Therefore, we expect 
that $\delta+\b_0$ takes values roughly between $-\pi/2$ and
$-(3/2)\pi$. 
This is confirmed by the numerical computation and the result is 
shown in Fig.1, where the allowed region of $x$ is taken to be 
$0.49<x<1$. The phase $\delta=\b_0$ varies roughly from $-\pi/2$ to 
 $-(2/3)\pi$ as we expected. 

\begin{figure}
\begin{center}
\includegraphics{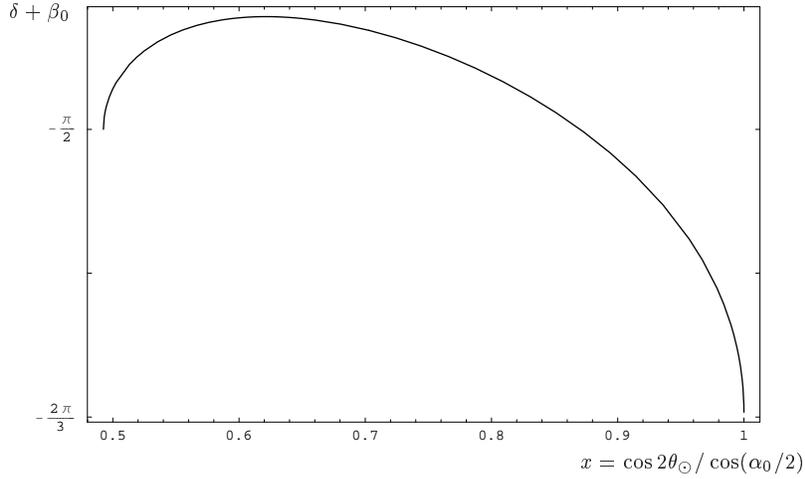}
\end{center}
\caption{The value of $\delta+\beta_0$ as a function of 
$x=\cos2\theta_{\odot}/\cos(\alpha_0/2)$. The region of $x$ 
is taken to be $0.49<x<1$.}
\end{figure}

For $\cos \a_0/2<0$, the $\delta$ is obtained just adding $\pi$ 
to values for $\cos \a_0/2>0$. 
The actual value of $\delta$ depends on $\beta_0$ which has to be 
fixed, maybe by the leptogenesis. 
\end{enumerate}

\section{The quantum corrections}
We consider the MSSM model associated with the right-handed 
neutrinos. That is, we assume the neutrino-Yukawa couplings 
and the heavy right-handed Majorana neutrino masses, and 
the neutrino mass matrix is generated by the see-saw mechanism. 

\begin{enumerate}
\item [(a)] The renormalization group effect
The renormalization group for the neutrino mass matrix in MSSM 
is given for $M_X>\mu>M_R$ by 
\bea
\frac{d m_\nu}{d \ln \mu}=\frac{1}{16\pi^2}\left\{ 
[(Y_\nu^\dagger Y_\nu)^T + (Y_e^\dagger Y_e)^T ]m_\nu
+m_\nu [(Y_\nu^\dagger Y_\nu) + (Y_e^\dagger Y_e)]
\right\},
\ena
where $Y_\nu$ and $Y_e$ are the Yukawa coupling matrices for 
neutrinos and charged leptons, respectively, aside from 
the terms proportional to the unit matrix. When $\mu<M_R$, 
we have to omit the neutrino-Yukawa contribution, because 
the heavy neutrinos decouple form the interaction. 

\begin{enumerate}
\item[(a-1)] The $\tau$-Yukawa coupling

In $Y_e$, the term which gives a mass to $\tau$ dominates and 
we take $(Y_e^\dagger Y_e)= 2{\rm diag}(0,0,(m_\tau/v_d)^2)$. 
In MSSM, $(m_\tau/v_d)^2=(m_\tau/v)^2(1+\tan^2\beta)$ and 
the $\tau$-Yukawa can give a sizable contribution for large 
$\tan \beta$ case. We note that in the standard model, there is no such 
enhancement factor and thus this contribution is negligible. 
In MSSM, we find the $\tau$-Yukawa contribution as
$\e_e=0$ and 
\bea
\e_\tau&=&\frac{1}{8\pi^2}(1+\tan^2\beta)(m_{\tau}/v)^2
\ln \left(\frac{m_Z}{M_X}\right) \nonumber\\
&=&-5.2\times 10^{-3}\left(\frac{1+\tan^2 \b}{1+15^2}\right)\;,
\ena
where we used $M_X=10^{16}$GeV, $m_Z=91.187$GeV, $m_\tau=1.777$GeV and 
$v=245.4$GeV. 
Since $-2\e_e+\e_\tau<0$, the solar mixing angle turns 
towards the dark side, which is unwanted for the maximal solar 
mixing case. This contribution becomes small if we take 
$\tan \beta$ around $5$. 

\item[(a-2)] The neutrino-Yukawa coupling

For the contribution due to the neutrino-Yukawa couplings, 
there is no $\tan \beta$ enhancement, so that we have to 
take relatively large $Y_\nu^\dagger Y_\nu$ of order 0.1. If we take 
the scenario that the slepton mixing occurs mainly from 
the renormalization group effect from $Y_\nu$, the 
lepton flavor violation process such as $\mu \to e+\gamma$ 
tells that the off diagonal terms of $(Y_\nu^\dagger Y_\nu)$ 
must be much smaller than $10^{-1}$. Therefore, it is reasonable 
to assume that $Y_\nu^\dagger Y_\nu$ is almost diagonal and 
thus we parameterize 
\bea
Y_\nu^\dagger Y_\nu&=&{\rm diag}(y_1^2,y_2^2,y_3^3)
\nonumber\\
&=&y_2^2 +{\rm diag}(y_1^2-y_2^2,0,y_3^2-y_2^2)\;.
\ena
Then, we find 
\bea
\e_i&=&\frac{1}{16\pi^2}(y_i^2-y_2^2)
\ln \left(\frac{M_R}{M_X}\right)\nonumber\\
&=&-4.4\times 10^{-3}\left(\frac{y_i^2-y_2^2}{0.1}\right)
\ln\left(\frac{M_R}{10^{12}{\rm GeV}}\right)\;,
\ena
where $i=e$, $\tau$ and $M_R$ is the mass of the 
heavy right-handed neutrinos. 
In order to get $\e_e<0$ and $\e_\tau>0$, we need 
the anti-hierarchical structure for $Y_\nu^\dagger Y_\nu$, i.e., 
$y_1^2-y_2^2>0$ and $y_3^2-y_2^2<0$ and the large 
coupling strength of order $0.3$ is required. 

\end{enumerate}

\item[(b)] The threshold corrections

There may be two kinds of threshold corrections. One is 
due to the mass splitting among heavy right-handed neutrinos, 
which was discussed by Antusch, Kersten, Lindner and Ratz \cite{nu_yukawa_2}
for the inverted Dirac neutrino mass case . 
The other is due to the slepton mass splitting \cite{slepton}. Here, we only
discuss 
the effect due to the slepton mass splitting. 

\begin{enumerate}
\item[(b-1)] The slepton mass splitting

We assume that the universal soft breaking terms aside from 
the corresponding term to the neutrino-Yukawa interaction which 
is non-universal. Since we assume that 
$Y_\nu^\dagger Y_\nu$ is diagonal, the off diagonal 
elements of the slepton mass matrix is not induced. The lepton 
flavor violation occurs from the slepton mass splitting. 
The effect to the neutrino mass matrix is given by \cite{slepton}
\bea
\epsilon_i=\frac{g^2}{32\pi^2}\left(-\frac1{z_i}
+\frac{z_i^2-1}{z_i^2}\ln (1-z_i)-(i\to \mu)  \right)
\ena
where $i=e, \tau$ and $z_i=1-(\tilde M_i/\tilde m)^2$, $\tilde m$ is 
wino mass and $\tilde M_i$ is the charged slepton mass. 

We find that $\e_i>0$ if $\tilde M_i>\tilde M_\mu\sim \tilde m$ and 
$\e_i<0$ if $\tilde M_i<\tilde M_\mu\sim \tilde m$
Thus, $\e_e<0$ and $\e_\tau>0$ are realized when the 
slepton mass splitting is hierarchical, i.e., 
$\tilde M_e<\tilde M_\mu<\tilde M_\tau$. 
In Fig.2, we showed values of $-2\e_e+\e_\tau$ as the function of 
$\tilde M_e/\tilde m$ and $\tilde M_\tau/\tilde m$ with 
$\tilde M_\mu=\tilde m$. Value of order $3\times 10^{-3}$ can be 
obtained for the ordinary mass ordered case, typically for 
$\tilde M_e/{\tilde m}\sim \tilde m/{\tilde M_\tau} \sim 1/2$.  
\end{enumerate}
\end{enumerate}

\begin{figure}
\begin{center}
\includegraphics{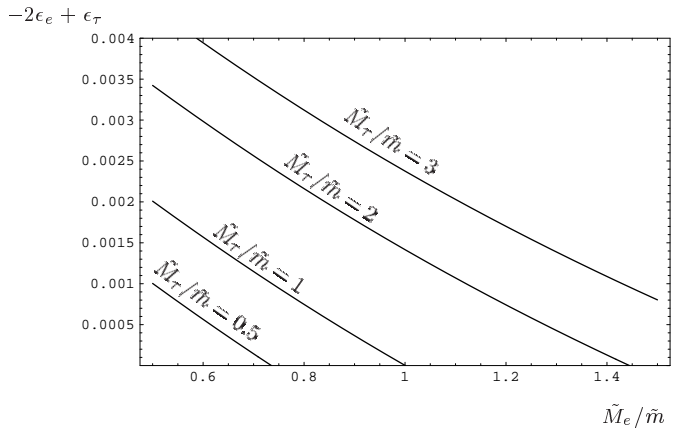}
\end{center}
\caption{The $\tilde{M}_e/\tilde{m}$ and $\tilde{M}_{\tau}/\tilde{m}$
dependence of the value of $-2\epsilon_e+\epsilon_{\tau}$ .
$x=\cos2\theta_{\odot}/\cos(\alpha_0/2)$. The region of $x$ 
is taken to be $0.49<x<1$.
In this figure, we take $\tilde M_{\mu}=\tilde m$ for simplicity.}
\end{figure}

\section{The summary}
We considered the fate of the bi-maximal mixing scheme which is 
realized at GUT in the light of the recent experimental implication 
that the solar neutrino mixing angles is in the normal side. 
We considered three kinds of quantum effects in the MSSM: 
a) The effect due to 
the  $\tau$-Yukawa coupling. We showed that this effect turns 
the solar angle towards the dark side as the energy scale 
decreases form the GUT scale to the weak scale. 
b)The effect due to the neutrino-Yukawa couplings $Y_\nu$. 
This contribution becomes effective when elements 
of $Y_\nu$ are as large as $0.3$, which gives the Dirac neutrino masses 
of order 50GeV. If we adopt the simple case where the soft breaking 
term is universal at the GUT scale. Then the only source for the 
the lepton flavor violation is from the neutrino-Yukawa couplings. 
In this story, the lepton flavor violation processes 
such as $\mu \to e+\gamma$ occurs from the renormalization group 
effect from $M_X$ to $M_R$ \cite{LFV-Ynu}, which is proportional to 
$Y_\nu^\dagger Y_\nu$. We argued that 
$Y_\nu^\dagger Y_\nu$ must be almost diagonal in order to suppress 
the lepton flavor violation process. 
In addition, the inverted hierarchy for the diagonal elements of 
$Y_\nu^\dagger Y_\nu$ is required, in order to obtain the normal side 
solar angle at low energy. c) We also considered the slepton threshold 
effect. We showed that the slepton threshold effect rotates the 
solar angle towards the normal side if the slepton mass spectrum 
is hierarchical. 

The bi-maximal mixing scheme which is achieved at GUT scale is 
quite interesting in the following reasons: 1)This mixing 
scheme can be obtained in the GUT scheme. 2)The neutrino mass 
matrix contains only seven parameters, three mixing angles, two 
Majorana phases and three masses. By using the renormalization group, 
the parameters at the low energy are obtained. In particular,  
the $|V_{13}|$ and the Dirac phase $\delta$ are induced by the 
renormalization group. Furthermore, if we assume that  
$Y_\nu^\dagger Y_\nu$ is almost diagonal, we can show that 
all CP violation phases are given by two Majorana phases, which 
are phases of neutrino masses. That is, we can show that 
the CP violation phases which appear in the leptogenesis are given by 
two Majorana phases, in the see-saw scheme. We gave this 
argument to demonstrate that the bi-maximal mixing provides the 
possibility that we understand CP violation phases. This work is 
now under preparation.

\acknowledgments{
This work is supported in part by 
the Japanese Grant-in-Aid for Scientific Research of
Ministry of Education, Science, Sports and Culture, 
No.12047218 and No.15540276.
The work of T.S. was supported in part by Research Fellowship of the
Japan Society for the Promotion of Science (JSPS) for
Young Scientists (No.15-03927).
}


\newpage
\begin{thebibliography}{99}
%%% Super-K-solar
\bibitem{SK-Solar}
S.~Fukuda {\it et al.}  [Super-Kamiokande Collaboration],
Phys.\ Rev.\ Lett.\  {\bf 86}, 5651 (2001).
%%% SNO-solar
\bibitem{SNO-Solar}
Q.~R.~Ahmad {\it et al.}  [SNO Collaboration],
Phys.\ Rev.\ Lett.\  {\bf 87}, 071301 (2001);
Q.~R.~Ahmad {\it et al.}  [SNO Collaboration],
Phys.\ Rev.\ Lett.\  {\bf 89}, 011302 (2002).
%%% KamLAND
\bibitem{KamLAND}
K.~Eguchi {\it et al.}  [KamLAND Collaboration],
Phys.\ Rev.\ Lett.\  {\bf 90}, 021802 (2003).
%%% Model of Bi-maximal
\bibitem{BiMax}
F.~Vissani,
arXiv:hep-ph/9708483;
V.~D.~Barger, S.~Pakvasa, T.~J.~Weiler and K.~Whisnant,
Phys.\ Lett.\ B {\bf 437}, 107 (1998);
A.~J.~Baltz, A.~S.~Goldhaber and M.~Goldhaber,
Phys.\ Rev.\ Lett.\  {\bf 81}, 5730 (1998).
%%% Previous paper
\bibitem{mst1}
T.~Miura, T.~Shindou and E.~Takasugi,
Phys.\ Rev.\ D {\bf 66}, 093002 (2002).
%%% Stability of neutrino mixing
\bibitem{Stability}
J.~A.~Casas, J.~R.~Espinosa, A.~Ibarra and I.~Navarro,
JHEP {\bf 9909}, 015 (1999);
%J.~A.~Casas, J.~R.~Espinosa, A.~Ibarra and I.~Navarro,
Nucl.\ Phys.\ B {\bf 573}, 652 (2000);
K.~R.~Balaji, A.~S.~Dighe, R.~N.~Mohapatra and M.~K.~Parida,
Phys.\ Rev.\ Lett.\  {\bf 84}, 5034 (2000);
N.~Haba, Y.~Matsui, N.~Okamura and T.~Suzuki,
Phys.\ Lett.\ B {\bf 489}, 184 (2000).
%%% Majorana phase
\bibitem{Majoranaphase}
S.~M.~Bilenky, J.~Hosek and S.~T.~Petcov,
Phys.\ Lett.\ B {\bf 94}, 495 (1980);
M.~Doi, T.~Kotani, H.~Nishiura, K.~Okuda and E.~Takasugi,
Phys.\ Lett.\ B {\bf 102}, 323 (1981);
J.~Schechter and J.~W.~Valle,
Phys.\ Rev.\ D {\bf 22}, 2227 (1980);
Phys.\ Rev.\ D {\bf 23}, 1666 (1981).
%%% Tau Yukawa 
\bibitem{tau_yukawa}
P.~H.~Chankowski and Z.~Pluciennik,
Phys.\ Lett.\ B {\bf 316}, 312 (1993);
K.~S.~Babu, C.~N.~Leung and J.~Pantaleone,
Phys.\ Lett.\ B {\bf 319}, 191 (1993);
N.~Haba and N.~Okamura,
Eur.\ Phys.\ J.\ C {\bf 14}, 347 (2000);
J.~R.~Ellis and S.~Lola,
Phys.\ Lett.\ B {\bf 458}, 310 (1999);
T.~Miura, E.~Takasugi and M.~Yoshimura,
Prog.\ Theor.\ Phys.\  {\bf 104}, 1173 (2000);
S.~Antusch, M.~Drees, J.~Kersten, M.~Lindner and M.~Ratz,
Phys.\ Lett.\ B {\bf 519}, 238 (2001);
Phys.\ Lett.\ B {\bf 525}, 130 (2002).
%%% Neutrino Yukawa 
\bibitem{nu_yukawa_1}
S.~Antusch, J.~Kersten, M.~Lindner and M.~Ratz,
Phys.\ Lett.\ B {\bf 538}, 87 (2002).
\bibitem{nu_yukawa_2}
S.~Antusch, J.~Kersten, M.~Lindner and M.~Ratz,
Phys.\ Lett.\ B {\bf 544}, 1 (2002).
%%% Slepton mass
\bibitem{slepton}
E.~J.~Chun and S.~Pokorski,
Phys.\ Rev.\ D {\bf 62}, 053001 (2000);
P.~H.~Chankowski, A.~Ioannisian, S.~Pokorski and J.~W.~Valle,
Phys.\ Rev.\ Lett.\  {\bf 86}, 3488 (2001)
\bibitem{mns}
Z.~Maki, M.~Nakagawa, and S.~Sakata,
Prog.\ Theor.\ Phys.\  {\bf 28}, 870 (1962).
%%% LFV and Nu-Yukawa
\bibitem{LFV-Ynu}
F.~Borzumati and A.~Masiero,
Phys.\ Rev.\ Lett.\  {\bf 57}, 961 (1986);
J.~Hisano, T.~Moroi, K.~Tobe, M.~Yamaguchi and T.~Yanagida,
Phys.\ Lett.\ B {\bf 357}, 579 (1995);
J.~Hisano, T.~Moroi, K.~Tobe and M.~Yamaguchi,
Phys.\ Rev.\ D {\bf 53}, 2442 (1996);
Phys.\ Lett.\ B {\bf 391}, 341 (1997);
J.~Hisano, D.~Nomura and T.~Yanagida,
{\it ibid.}\ {\bf 437}, 351 (1998);
J.~Hisano and D.~Nomura,
Phys.\ Rev.\ D {\bf 59}, 116005 (1999);
W.~Buchm\"{u}ller, D.~Delepine and F.~Vissani,
Phys.\ Lett.\ B {\bf 459}, 171 (1999);
W.~Buchm\"{u}ller, D.~Delepine and L.~T.~Handoko,
Nucl.\ Phys.\ B {\bf 576}, 445 (2000);
J.~R.~Ellis, M.~E.~Gomez, G.~K.~Leontaris, S.~Lola and D.~V.~Nanopoulos,
Eur.\ Phys.\ J.\ C {\bf 14}, 319 (2000);
J.~Sato and K.~Tobe,
Phys.\ Rev.\ D {\bf 63}, 116010 (2001);
J.~Sato, K.~Tobe and T.~Yanagida,
Phys.\ Lett.\ B {\bf 498}, 189 (2001);
S.~Lavignac, I.~Masina and C.~A.~Savoy,
J.~A.~Casas and A.~Ibarra,
Nucl.\ Phys.\ B {\bf 618}, 171 (2001);
{\it ibid.}\ {\bf 520}, 269 (2001);
Nucl.\ Phys.\ B {\bf 633}, 139 (2002);
F.~Deppisch, H.~Pas, A.~Redelbach, R.~R\"{u}ckl and Y.~Shimizu,
arXiv:hep-ph/0206122;
J.~R.~Ellis, J.~Hisano, M.~Raidal and Y.~Shimizu,
Phys.\ Rev.\ D {\bf 66}, 115013 (2002);
S.~Pascoli, S.~T.~Petcov and C.~E.~Yaguna,
arXiv:hep-ph/0301095.
S.~T.~Petcov, S.~Profumo, Y.~Takanishi and C.~E.~Yaguna,
arXiv:hep-ph/0306195.
\end{thebibliography}
\end{document}